# Eight-channel reconfigurable microring filters with tunable frequency, extinction ratio and bandwidth

Hao Shen,[1,2,*] Maroof H. Khan,[1,3] Li Fan,[1,2] Lin Zhao,[1,2] Yi Xuan,[1,2] Jing Ouyang,[1,2] Leo T. Varghese, [1,2] and Minghao Qi[1,2,*]

[1]*Birck Nanotechnology Center, Purdue University, 1205 W. State Street, West Lafayette, Indiana, 47907, USA*
[2]*School of Electrical and Computer Engineering, Purdue University, 465 Northwestern Avenue, West Lafayette, Indiana, 47907, USA*
[3] *Department of Physics, Purdue University, 525 Northwestern Avenue, West Lafayette, Indiana, 47907, USA*
[*]*shen17@purdue.edu,* [*]*mqi@purdue.edu*

**Abstract:** We demonstrate an eight-channel reconfigurable optical filter on a silicon chip consisting of cascaded microring resonators and integrated compact heaters. With an embedded Mach-Zehnder (MZ) arm coupling to a microring resonator, the important parameters of a filter such as center frequency, extinction ratio and bandwidth can be controlled simultaneously for purposes of filtering, routing and spectral shaping, thus making our method potentially useful in dense wavelength division multiplexing (DWDM) and radio frequency arbitrary waveform generation (RFAWG). Multichannel filter response was successfully tuned to match the International Telecommunication Unit (ITU) grid with 50, 100 and 200GHz in channel spacing. Programmable channel selectivity was demonstrated by heating the MZ arm, and continuous adjustment of through-port extinction ratio from 0dB to 27dB was achieved. Meanwhile, the 3dB bandwidth in the drop port changed from 0.11nm to 0.15nm due to the heating. The device had an ultra-compact footprint (1200μm×100μm) excluding the metal leads and contact pads, making it suitable for large scale integration.

---

**1. Introduction:**

Silicon-on-insulator (SOI) microphotonics is an attractive technology to shrink photonic systems down to micrometer-scale size, and the miniaturization will significantly reduce the size and power consumption of functional modules in optical communication system as well as enable novel applications[1-3]. Micro-resonators are among the most fundamental components and have been demonstrated to make commercial-grade reconfigurable filters[4].

The process compatibility to standard complementary metal oxide semiconductor (CMOS) allows one to design on-chip systems with both electronic circuits and optical modules. The extremely small size (micrometer scale) and simple architecture of micro-resonators are especially suitable for multiple channel applications such as dense wavelength-division-multiplexing (DWDM). The majority of other explored functionalities such as modulators[1, 5], optical add-drop multiplexers[6] and optical switchers[7] also require precise working frequency and full programmability for both long-distance signal transmission and chip-scale optical interconnection.

Unfortunately, SOI micro-resonators are extremely sensitive to fabrication variations and imperfections. Very small size variations, even a few picometers, can cause appreciable resonance frequency shifts[8]. Therefore, achieving the designed absolute resonance frequency of a micro-resonator through open-loop control of the fabrication process is difficult. The post-fabrication trimming method[9] may be one solution. However, it is open-loop and is restricted by the availability of suitable materials and trimming range. A simpler and more controllable solution, such as continuous tunability, is preferred to mitigate the fabrication uncertainty. Meanwhile, in many applications the continuous tunability, or programmability, is crucial for flexibility in data transmission, switching and routing. Fortunately, silicon has a relatively large thermo-optic coefficient and is thus amenable to thermal tuning. For the same reason, the micro-environment in which individual resonators reside needs to be temperature controlled for wavelength stability. Compared to other tuning schemes, such as current injection or laser illumination, thermal tuning provides larger range of control[10] and the devices (micro-heaters) are also easier to fabricate. The speed of thermal tuning used to be low, in the millisecond range. However, recently it has been improved to microsecond range by direct heating method[11].

For devices like microring add-drop filters, the parameters to be tuned include not only resonance frequency on which most of the previous reports focused, but also the 3dB bandwidth in drop port, and extinction ratio in some novel applications, such as spectral shaper for radio frequency arbitrary waveform generation[12]. With heaters fabricated above each microring and the embedded Mach-Zehnder (MZ) arm, we can individually and simultaneously control the center frequency, extinction ratio and bandwidth of every channel in a multi-channel filter[12].

In this paper we present our work on the accurate control of various parameters in a multichannel microring filter. First a design with straight through-port waveguide and dissipated drop ports was fabricated to demonstrate the fabrication process and thermal control capability. In the following section we introduced an eight-channel filter with through port combined with MZ arms and a common drop port. The filter spectral responses fitting the ITU grid with 50, 100 and 200GHz channel spacing were then achieved with the same device, demonstrating the wide, multi-channel tunability. The control of extinction ratio and bandwidth was then verified, and on-off channel selection control, or on-off keying, was realized. With the periodic spectral responses of microring resonators, which are spaced by a Free Spectral Range (FSR), the filter can cover a large band of spectrum.

## 2. Eight-channel filter coupled to a straight waveguide

The first demonstration is a traditional design with eight microring resonators coupled to a straight through-port waveguide. Micro-ring resonators were fabricated on an SOI wafer which has a 3μm buried silicon dioxide layer and 250nm thick top silicon layer. The buried oxide layer mitigates the optical power leakage into the substrate. The ultra-compact cross-section (500nm×250nm) of photonic wire strongly confines the field in the waveguide area. Thus it reduces the scattering loss due to the side-wall roughness on the waveguide. For the same reason the bending loss is also very low, almost negligible when the bending radius is larger than 5μm. As those two major types of loss are suppressed, the round trip loss in microring resonator is reduced, which yields very high intrinsic quality factor (Q value). We have already demonstrated intrinsic quality factors up to 270,000 when the radius of microring was 5μm, which corresponded to ~3.5dB/cm waveguide propagation loss[13].

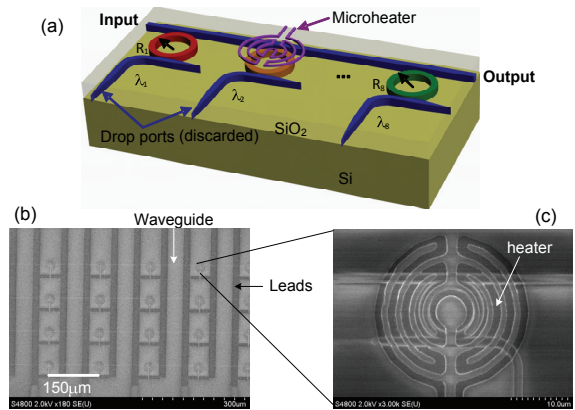

Fig.1: (a) Schematic of the 8-channel resonator array. (b) A scanning electron micrograph of the fabricated micro heater array, with high packing density. Four sets of devices were stacked vertically. (c) A zoom-in view of the micro heater over a microring.

In Fig.1a, the eight microring resonators had slightly different radii so that each ring resonated at different wavelengths. The radius of the first microring resonator in the array was around 5μm, with an FSR ~16nm. The radius of each subsequent ring increased sequentially by 8nm, and yielded approximately a 1nm increment in resonance wavelengths. The drop ports were all dissipated. The coupling gap between the microring resonator and the straight waveguide was 200nm, which is in the over-coupling region. We fabricated spiral shaped titanium heaters (Fig. 1c) above each ring to confine the heat locally to the waveguide. Each ring was separated by 150μm (Fig. 1b) in order to confine the heat locally and reduce thermal cross talk. Silicon has a relatively large thermo-optic coefficient ($\partial n/\partial T = 1.86 \times 10^{-4} K^{-1}$), which makes thermal tuning a preferred method when the resonance wavelength shift needs to be several nanometers or more. Tuning speed used to be the bottleneck for thermal tuning. The heaters generally sit on a relatively thick buffer layer (*e.g.* > 1μm thick $SiO_2$) to avoid excessive optical loss, and the small thermal conductance of $SiO_2$ limits the tuning speed. Recently the tuning speed was improved to microsecond level by placing the electrical contacts on the silicon resonator directly to heat silicon, without any oxide buffer layer[14].

Bus waveguide with 500nm width and microring waveguide with 600nm width were written with a 100kV e-beam lithography tool (Vistec VB6) on 200nm of HSQ negative resist. Subsequent etching was done with HSQ as a mask, in a plasma of chlorine and argon mixture in a Panasonic inductive coupled plasma etching tool. The dry etching was tuned to yield a sidewall roughness below 10nm, which significantly reduced the sidewall scattering loss. The chip was then dipped in hydrofluoric acid to remove the remaining HSQ mask, followed by the deposition of 600nm silicon dioxide through plasma enhanced chemical vapor deposition (PECVD). Since PECVD deposition is not completely conformal, the step-shaped topography of the silicon waveguides remained on the top surface of the oxide, and could lead to broken metal wires when metal is evaporated over such steps. To smooth out the aforementioned features, 600nm of HSQ was spun over the PECVD oxide. This allowed continuous films of titanium and gold to be deposited for the heater and connecting wires. Titanium strip width was 1 μm and its thickness was 150nm. The spiral-shaped heater had a resistance of 1.2kΩ. Gold lines connecting the common ground pads were patterned with optical lithography, whereas for Ti heaters e-beam lithography was once more applied due to stringent alignment requirements.

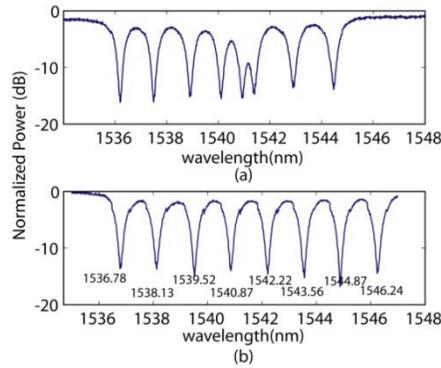

Fig. 2. Through-port power spectrum of an eight channel microring filter.
(a) before thermal tuning, (b) after thermal tuning.

Fig. 2a shows the normalized power spectrum of our eight-channel filter without thermal tuning. We designed the radii of the rings to achieve identical channel spacing. However, in fig. 2a it is clear the resonance wavelengths were unevenly distributed with large standard deviations. This deviation mainly comes from the limited precision of the digital-to-analog converter and internal distortion of the exposure field in electron-beam lithography. A group in MIT has controlled the fabrication process carefully to compensate those effect[15]. Here we demonstrate another solution with thermal tuning, which red-shifts the resonance wavelengths of microring resonators. Before thermal tuning, the filter array had a resonance wavelength spacing averaged at 1.18nm, but with standard deviation of 0.39nm. After careful thermal tuning, as shown in Fig. 2b, the 8 channel filter response had an average dip spacing of 1.35nm with standard deviation of only 0.025nm. In each channel, the through port extinction ratio was larger than 15dB.

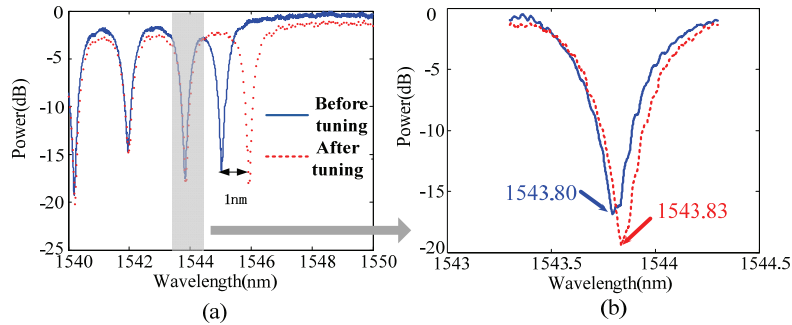

Fig.3: (a) Thermally tuning the resonance wavelength of the microring with the largest radius by 1nm. The three neighboring resonators are also shown. (b) The resonance wavelength of the adjacent channel is shifted by 0.03nm due to thermal crosstalk.

In our device, the 3μm silicon dioxide layer functions as a heat insulator to reduce the heat conduction from one microring resonator to another. The silicon substrate layer underneath the 3μm oxide layer is, however, a good thermal conductor. Here we investigate the thermal crosstalk (Fig. 3). Only the microring with the largest diameter was heated and its resonance frequency was red-shifted by 1nm. The heat crosstalk will affect the resonance frequencies of other channels on this chip. Checking the adjacent channel, the resonance wavelength was shifted by 0.03nm after the thermal tuning. Thus we roughly claim our thermal crosstalk to be 3%.

Compared to silica devices such as arrayed waveguide grating (AWG), silicon microring filters generally consume less power for thermal control and tuning[16]. In our design, 5mW power was needed to red shift the resonance wavelength of a microring resonator by 1nm. In order to reduce the thermal crosstalk, the sample sat on an aluminum block functioning as a heat sink. Without the aluminum heat sink, the thermal crosstalk increased to 10%. Additional cooling mechanism can be applied to further decrease the thermal crosstalk.

## 3. Eight channel filter with embedded Mach-Zehnder Arm

Resonance frequency is not the only tunable parameter in microring based filters. The through port extinction ratio and 3dB bandwidth are also important parameters that might require tuning in filter design and optical waveform generation[17]. For example, in optical waveform generation using the through port signal, the extinction ratios determines the individual valleys of a waveform, and the 3dB bandwidths of individual resonances determine the duty cycle of the waveform. One way to achieve such tunability is to modify the straight through port waveguide to an arc-shaped arm, which has two coupling regions to the resonator (Fig. 4a). This structure can be viewed as an embedded MZ arm on each microring resonator. It has been demonstrated that this interference arm can effectively change the coupling coefficient[4, 18, 19] between the through port and resonator, and consequently change the through-port attenuation and drop-port loss. However, applying such a scheme to multiple channels has not been demonstrated. Here we demonstrate an advanced design of the all-tunable 8-channel filter (Fig. 4).

Additionally, instead of dissipating the drop ports, here we add a common drop port to download all the eight channels together to one output port. This is a simple solution to download multiple channels at the same time to a certain port, without suffering the loss from the Y-branch combiners[20]. The multi-cavity effect does exist in between the multiple microring resonators. However since in our application the resonance frequency of each channel is separated, the interference is limited.

The microring geometry and fabrication process was similar to the previous design. The MZ arm consisted of two sections. The arc arm had a radius of 15μm for the benefits of large heating area. Two transition arc (5μm in radius) waveguides smoothly connected the MZ arm and the straight waveguide. The gaps between the MZ arm and microring resonator were both 300nm. In this device the heater was still fabricated from titanium but the contact leads were made from aluminum.

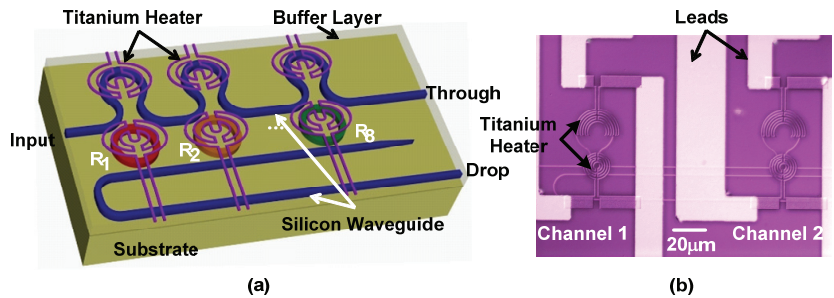

Fig.4: (a) Schematic of a reconfigurable eight-channel microring filter with MZ arms. (b) Optical microscopic picture of the fabricated device.

The MZ arm changes the through-port coupling coefficient, which enables the manipulation of extinction ratio and bandwidth in the through port power spectrum. Consequently, the peak intensity in drop port spectrum will change according to the changes in the through port extinction. Fig. 5a shows the continuous control of extinction ratio from minimum (0dB) to maximum (27dB), when we tuned the heater on the MZ arm of the channel at 1552nm. The resonance wavelength and extinction ratio of the adjacent channel did not show obvious shift, which reassures the low crosstalk. Fig. 5b shows the voltage required on the heater for tuning the through port extinction ratio. From the initial 5.67dB to the maximum 27dB through-port extinction ratio, we needed to apply ~20mW power (10.3V, 5kΩ resistance for the heater) on the MZ arm. The through-port 3dB bandwidth varied from 0.076 to 0.163nm. The drop-port 3dB bandwidth varied from 0.117 to 0.162nm. The extinction ratio dropped from 27dB back to 0dB when more heat was applied to the MZ arm. The extinction ratio of the as-fabricated resonator, when no voltage was applied to the heater, depended on the exact dimension of the MZ arm and ring. Due to the fabrication variations, the as-fabricated extinction ratio is generally unpredictable. A tuning element, in this case a micro-heater, is always necessary to achieve desired coupling coefficient.

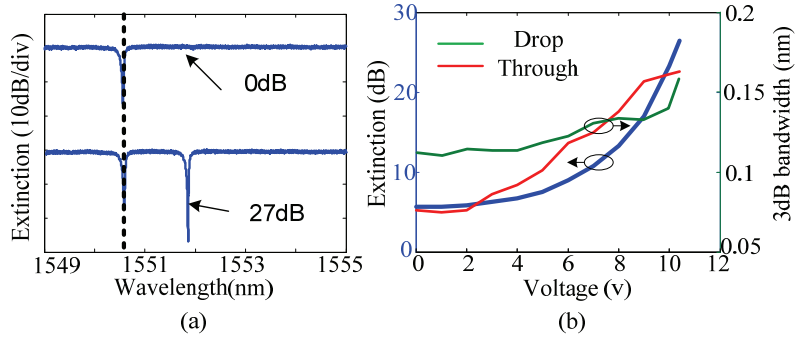

Fig.5 (a) Continuous control of through-port extinction ratio from 0dB to 27dB by heating the MZ arm coupled to a microring resonator. (b) The through-port extinction ratio, through-port 3dB bandwidth and drop-port 3dB bandwidth vs. voltage applied to the heater over the MZ arm.

Fig. 5b shows that the 3dB bandwidth changed together with extinction ratio. The reason for this connection is we only have one MZ arm at the through port. The through port extinction ratio and drop port 3dB bandwidth can be described as:

$$T_{through} = \frac{\left(\kappa_d^2 + \kappa_l^2 - \kappa_e^2\right)^2}{\left(\kappa_d^2 + \kappa_l^2 + \kappa_e^2\right)^2} \quad \text{Eq.1a}$$

$$\delta_{3dB-drop} = \frac{FSR}{2\pi}\left(\kappa_d^2 + \kappa_l^2 + \kappa_e^2\right)^2 \quad \text{Eq.1b}$$

where FSR is the free spectral range, $\kappa_d$, $\kappa_l$ and $\kappa_e$ are the drop port coupling coefficient, micro-cavity round trip loss and through port coupling coefficient, respectively [21]. In our particular case, our as-fabricated (without heating) MZ arm was in the under-coupling regime. With increasing heat, the $\kappa_e$ kept increasing until it reached the critical coupling condition, $\kappa_d^2 + \kappa_l^2 = \kappa_e^2$. From Eq. 1b, the drop port 3dB bandwidth increased with the extinction ratio (Fig. 5b). Adding MZ arm on both the through port and drop port is one solution to control the extinction ratio and 3dB bandwidth independently [19]. In our device, the current was fed to the heater by direct contact with tungsten probes. Due to the limited number of probes (16 in our case), only two heaters could be allocated for each channel (a total of 8 channels) so we could not add another MZ arm on the drop port with integrated heater control. Direct probe contact will always have the limitation on the available number of probes. Wire bonding [22] is the choice if we need to apply more heaters in the future.

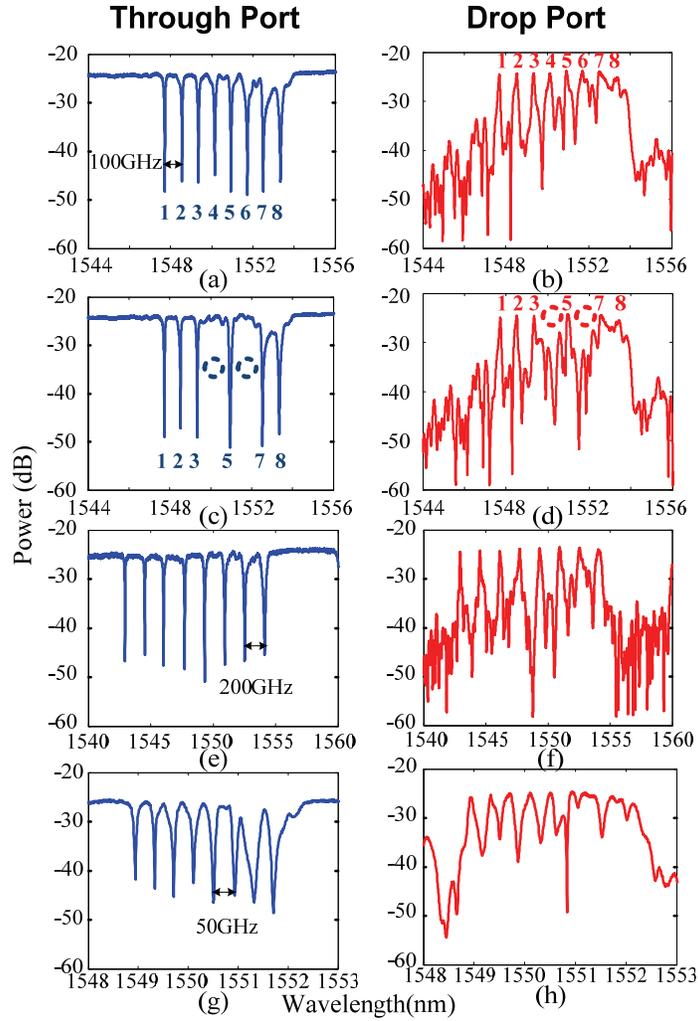

Fig 6. Eight channel microring filter with channel frequency from (a-b):193.0 to 193.7THz (100 GHz channel spacing); (c-d): 193.0 to 193.7THz (100 GHz channel spacing) with channel 4 and 6 suppressed; (e-f): 192.9 to 194.3 THz (200 GHz channel spacing); (g-h):193.20 to 193.55THz (50 GHz channel spacing). The blue charts are transmission spectra from through port and the red ones are transmission spectra from common drop port.

With the control elements on every resonator and MZ arm, we can build the multichannel filters fitting the ITU grid precisely, and balance the through-port extinction ratio at the same time. Fig. 6a demonstrates the eight channel drop filter from 193.0 to 193.7THz with exactly 100GHz channel spacing. All the extinction ratios were tuned to approach the largest value (~25dB). In Fig. 6b the drop port shows corresponding multi-channel peaks. In this case, the tuning range of resonance frequency for microring resonator was more than one FSR (16nm). Thus with appropriate filter, the working band of this filter can cover multiple communication bands. As shown in Fig. 6b, there were additional peaks in the drop port, in-between two channels (*e. g.* between channels 4 and 5), and we suspect that they come from the multi-cavity resonance effect induced by the common drop port. The suppression of this peak was 9dB (outside the band) since the resonance wavelengths of different microring cavities were relatively far away. We notice that the qualify factors of channels 6, 7, and 8 were significantly degraded, mainly due to the large amount of heat applied to the rings and their associated MZ interference arms, which may also contributed to the Fano resonance [23]. The degraded line width and extra resonance in these channels may render them unsuitable for practical applications and further investigation is being conducted to mitigate their impacts.

In Fig. 6c-6d channel 4 at 193.4THz and channel 6 at 193.2THz were tuned to the bypass state. The power in these two frequencies goes to the through port instead of drop port. This function will benefit the wave band routing [24] and hitless switching [25]. In Fig. 6e-6f and 6g-6h) we show the filter response with 200GHz (192.9 to 194.3THz) and 50GHz (193.20 to 193.55THz) channel spacing, respectively. When the channel spacing was small (50GHz), the inter-channel crosstalk was severe since the 3dB bandwidth of the single microring cavity is close to the channel spacing (Fig. 6g-6h). Finally, the fiber to fiber loss could be up to 25dB. Around 10dB/facet loss came from the mode mismatch between the compact silicon waveguide to the tapered lensed fiber. This loss can be reduced by applying inverse taper [26] or grating coupler [27].

## 4. Discussion:

As we mentioned before, due to the limited control in fabrication process, the resonance wavelengths in as-fabricated multichannel filters cannot be evenly distributed with 1nm spacing. Moreover, the small length difference between MZ arms could lead to significant reduction in extinction ratio in the optical spectrum when no thermal tuning was applied (Fig. 7). In applications such as spectral shaping [12], extinction ratio of every channel is desired to reach the maximum. Thus the power consumption in heating for each channel could be quite different. It has been reported that uneven temperature distribution in ultra-compact silicon waveguides could cause thermal optical polarization rotation [28]. Depending on the different operating temperatures in each channel, the polarization might no longer be identical in each channel, especially for the channels at longer wavelength, where generally more heat are applied. That could be the reason for the relatively low quality factors and strong Fano effect in specific channels (No.7 and 8 in Fig. 6) in our device.

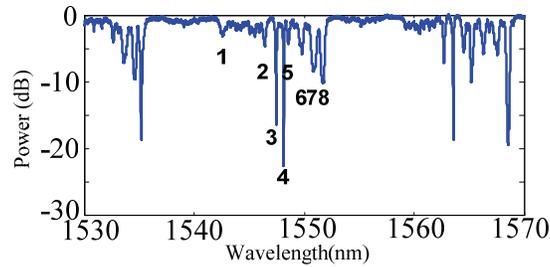

Fig.7: Through port power spectrum of eight channel microring resonator filter with embedded MZ arm before applying thermal tuning. The numbers indicate the channel positions.

Thermal drift comes from the environmental variation and the fluctuation of the DC source. The environmental temperature fluctuation can be mitigated by using a thermoelectric cooler. The DC voltage sources we used have a resolution of 10mV. With the estimation using the data in section 3, the average power drift was 0.04mV (at ~10V, 5kΩ). Thus the drift of resonance frequency in each channel was as low as 8 picometers, which is acceptable here. It can be further optimized by improving the stability of DC source.

The periodical spectral response of microring resonators benefits this device since our device can work in many different frequency bands by selecting different free spectral ranges (FSRs). Meanwhile, due to the waveguide dispersion, our tuning cannot achieve the precise spacing in resonance wavelengths in all free spectral ranges. Therefore, the channels selected are all confined to within one single FSR, and the extra channels in other FSRs may interfere with other channels intended to work in those frequency bands. Vernier effect [29] can be applied to eliminate the periodic responses of a single microring resonator by coupling two resonators with different radii in one channel. The channel survives only when the two resonators are tuned to the same resonance wavelengths, and the periodic spectral responses are eliminated due to the different FSRs of the two rings.

## 5. Conclusion:

In this paper we fabricated an eight-channel optical notch filter by cascading eight different microring resonators to a bus waveguide. With compact heaters fabricated on each microring, we demonstrated a reconfigurable filter with controllable channel frequency spacing. Moreover, an advance strategy which couples the microring resonator to an embedded MZ arm was designed and fabricated. Programmable filter responses precisely matching ITU grid were realized. Multi-dimensional control in frequency, extinction ratio and bandwidth with our filter design was demonstrated, which makes this design potentially suitable for the optical communication and photonic microwave applications.

**Acknowledgement:**
The work was supported by the National Science Foundation under contract ECCS-0701448, and by the Defense Threat Reduction Agency under contract HDTRA1-07-C-0042.